\documentstyle[twoside,fleqn,espcrc2,epsf]{article}

\newcommand{\wt}{\widetilde}

\newcommand{\NN}{{\cal N}}

\newcommand{\be}{\begin{equation}}
\newcommand{\ee}{\end{equation}}
\newcommand{\ben}{\begin{eqnarray}\displaystyle}
\newcommand{\een}{\end{eqnarray}}
\newcommand{\refb}[1]{(\ref{#1})}
\newcommand{\sectiono}[1]{\section{#1}\setcounter{equation}{0}}

\title{Recent Developments in Superstring Theory\hfill \rm 
\small hep-lat/0011073}

\author{Ashoke Sen
\address{ Harish-Chandra Research Institute \\
(Formerly
Mehta Research Institute of Mathematics and Mathematical Physics) \\
Chhatnag Road, Jhusi, Allahabad 211019, INDIA
}
\thanks{E-mail: sen@mri.ernet.in, asen@thwgs.cern.ch}
}

\begin{document}

\begin{abstract}
The talk contains a brief introduction string theory, 
followed by a discussion of some of the recent
developments.
\end{abstract}

\maketitle

This talk will begin with a brief review of pre-1994 string theory. Then I
shall describe the develoments during 1994-1996 which include duality
symmetries, D-branes and black holes. Finally I shall turn to the
discussion of some of the developments during 1997-2000. The topics
covered will include (M)atrix theory, Maldacena conjecture,
non-commutative
geometry, tachyon condensation, special limits of string theory in which
gravity decouples (little string theories, OM theory etc.) and large
radius compactification. This however is not an exhaustive review of
string theory. Indeed, many of the important developments in this subject,
like mirror symmetry, new ideas about cosmological constant, etc. will be
left out of this review. I shall work in the convention $\hbar=1, c=1$. 

\sectiono{String theory: pre-1994} \label{s1}

\begin{figure}
\begin{center}
\leavevmode
\epsfbox{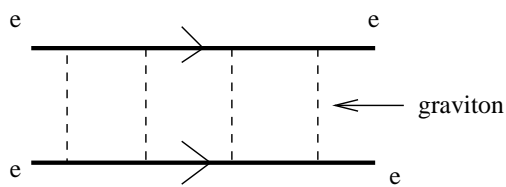}
\end{center}
\caption{Ultraviolet divergent graphs in quantum gravity.} \label{f1}
\end{figure}

As we all know, 
quantum field theory has been extremely successful in providing a
description of elementary particles and their interactions.
However, it does not work so well for gravity. The reason for this is
ultraviolet divergences. In particular Feynman graphs involving
gravitons, of the type shown in Fig.\ref{f1}, give divergent answers which
cannot be made finite by
standard renormalization procedure.
String theory is an attempt to solve this problem.

\begin{figure}
\begin{center}
\leavevmode
\epsfbox{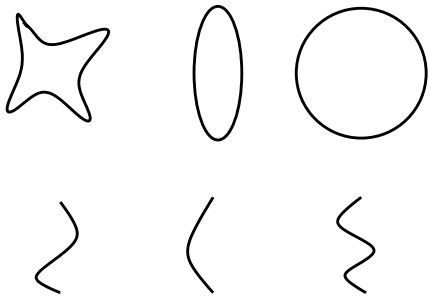}
\end{center}
\caption{Different vibrational modes of closed and open strings.}
\label{f2}
\end{figure}

The basic idea in string theory is quite simple\cite{GSW,POLCHINSKI}.
According to this theory, 
different elementary particles which we observe in nature are different
vibrational states of
a one dimensional object, {\it i.e.} a string. In other words, instead of
having different kinds of elementary
particles, we have only one kind of string, and the differences in the
observed properties of elementary particles arise because they correspond
to different quantum states of this string. Strings can be closed with no
ends
or open with suitable boundary condition at the two ends, as shown in
Fig.\ref{f2}. 

At the first sight, this would seem to contradict
what we know about elementary particles like electrons and quarks: they
behave like particles rather than one dimensional objects. However, when
one estimates the typical size of a string, one finds that this size is of
the order of $ 10^{-33}$cm $\sim \,\,  (10^{19}GeV)^{-1}$.\footnote{This 
size is controlled by the energy per unit length of the string, known as 
the string tension, which, in turn, is determined from the observed
value of the Newton's constant of gravitation. As we shall see later,
large radius compactification
can drastically change the estimate of the string size, but it is still
smaller than the distance scale that can be probed by current
accelerators.} 
This is much
smaller than the distance scale which can be probed by present day
accelerators. Thus there is no immediate contradiction with
experiments, since strings of such a small size will appear point-like in
present day experiments.

There is a well defined procedure for quantizing a string without
violating
Lorentz invariance.
It turns out that
\begin{itemize}
\item Quantum string theories do not suffer from any ultraviolet
divergence.
\item Spectrum of a string theory contains a particle which has
all the properties of a graviton $-$ the mediator of gravitational
interactions.
\end{itemize}
Thus string theory gives a finite quantum theory of gravity +
other stuff.

There are however several problems with string theory. They are as
follows:
\begin{itemize}
\item String theory is consistent only in (9+1) dimensional
space-time instead of the (3+1) dimensional space-time in which we seem
to live.

\item Instead of a single consistent string theory, there are {\it five}
consistent string theories in (9+1) dimensions. They are called
type IIA, type IIB, type I, 
SO(32) heterotic, and $E_8\times E_8$ heterotic string theories. On the
other hand it is desirable that we have a single theory, as there is only
one nature which string theory attempts to describe.
\end{itemize}
It turns out that
the first problem is resolved via a procedure called
compactification.
The second problem is partially resolved due to a property of string 
theory called duality; this will be discussed in the next section. 

\begin{figure}
\begin{center}
\leavevmode
\epsfbox{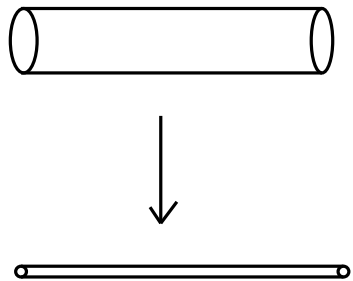}
\end{center}
\caption{Compactification of a two dimensional theory to a one dimensional
theory.} \label{f3}
\end{figure}

I shall now briefly discuss the idea of compactification.  The idea in
fact is quite simple:  take 6 of the 9 spatial directions to be small and
compact.  As long as the sizes of the compact directions are smaller than
the reach of the present day accelerators, the world will appear to be
(3+1) dimensional.  Since it is hard to draw a 9-dimensional space, a
caricature of compactification has been shown in Fig.\ref{f3} which
demonstrates how a 2-dimensional space can be made to look like a one
dimensional space.  Here we take the two dimensional space to be the
surface of a cylinder of radius $R$ and infinite length. For large $R$
(larger than the range of the most powerful telescope) the space will look
like an ordinary two dimensional space of infinite extent in all
directions. whereas for small $R$ (smaller than the resolution of the most
powerful microscope)  the space will look one dimensional. This way an
intrinsically two dimensional space can be made to look one dimensional.
The same idea works in making an intrinsically 9 dimensional space look 3
dimensional. 

It turns out that not all six dimensional compact manifolds can be
used for this purpose, but there are many choices. 
A particularly
important class of six dimensional manifolds, which can be used for
string compactification, are known as Calabi-Yau manifolds. 
Thus each string theory is (9+1) dimensions gives rise to many
different string theories in (3+1) dimensions after
compactification.
Some of these theories come tantalizingly close to the observed
universe\cite{CHSE}. In particular, they have:
\begin{itemize}
\item Gauge group containing $SU(3)\times SU(2)\times U(1)$
\item Chiral fermions
\item Three generations of quarks and leptons
\item N=1 supersymmetry
\end{itemize}
etc. However, so far nobody has found a suitable compactification
scheme which gives results in complete quantitative agreement with the
observed universe (including masses of various elementary particles).

\sectiono{String Theory: 1994-1996} \label{s2}

\subsection{Duality in String Theory} \label{s2.1}

Existence of duality symmetries in string theory started out as a
conjecture and still remains a conjecture. However so many
non-trivial tests of these conjectures have been performed by now
that most people in the field are convinced of the validity of   
these conjectures\cite{FILQ,9402002,9411178,9410167,9503124,9505105}. 

A duality conjecture is a statement of equivalence between two or
more apparently different string theories. Two of the most
important features of a duality relation are as follows:
\begin{itemize}
\item Under a duality map, often an elementary particle in one
string theory gets mapped to a composite particle in a dual
string theory and vice versa.
Thus classification of particles into elementary and composite
loses significance as it depends on which particular theory we
use to describe the system.
\item Duality often relates a weakly coupled theory to a strongly
coupled theory. If we denote by $g$ and $\wt g$ the coupling constants in
the two theories related by the duality map, then often there is a simple
relation between them of the form:
\be \label{e2.1}
g = (\widetilde g)^{-1}\, .
\ee
Thus a perturbation expansion in $g$ contains information about
non-perturbative effects in $\widetilde g$ in the dual theory. In
particular,
tree level (classical) results in one theory 
is given by the
sum of perturbative and non-perturbative
results in the dual theory.
\end{itemize}
{}From this it should be clear that duality is a property of the quantum
string theory and not its classical limit.

I shall now give some examples of duality:
\begin{itemize}
\item SO(32) heterotic string theory has been found to be dual to type I
string theory in (9+1) dimensions.
\item SO(32) (as well as $E_8\times E_8$) heterotic string theory
compactified on a
four dimensional
torus
$T^4$
has been found to be dual to 
type IIA string theory compactified on a non-trivial four
dimensional compact manifold called $K3$.

\item Type IIB string theory has been found to be self dual, {\it i.e.}
the theory at
coupling constant $g$ has been found to be equivalent to the same theory
with
coupling constant $1/g$.

\item The SO(32) (as well as $E_8\times E_8$) heterotic string theory,
compactified
on a six dimensional torus $T^6$, has been found to be self-dual in the
same sense.
\end{itemize}

\begin{figure}
\begin{center}
\leavevmode
\epsfbox{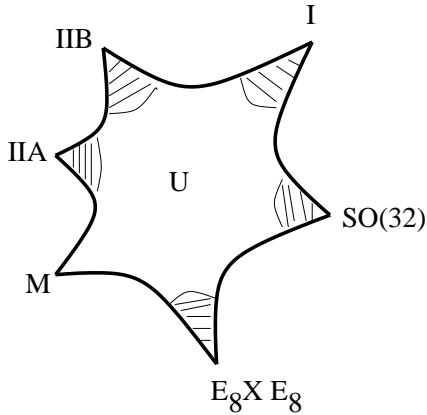}
\end{center}
\caption{The parameter space of U/M theory.} \label{f5}
\end{figure}

Since
duality relates different compactifications of different string theories,
it provides us with a unified picture of all string theories. According to
our
new understanding based on duality, there is a single underlying theory
$-$ which I shall call the $U$-theory $-$ with many degenerate vacua
labelled by a set of
parameters. Fig. \ref{f5} gives a schematic representation of this
parameter space. In some special limits, the $U$-theory can be described
by one of the (compactified) weakly coupled string theories.  These
special regions have
been shaded in the figure.  The rest of the regions represent string
theories at finite / strong coupling.  Note that besides the five corners
labelling five weakly coupled string theories, there is another corner
called $M$. This turns out to be a (10+1) dimensional theory, whose low
energy limit is the (10+1) dimensional supergravity
theory\cite{9501068,9503124,9607201}. Often people
use the same symbol $M$ to describe this special corner as well as the
whole
region of the parameter space, but I shall reserve the name $M$ for this
special corner, and use the symbol $U$ for the underlying theory with 
the full parameter space.

Thus we see that using the idea of duality, we can combine all string
theories into a single theory, with a big parameter space. It turns out
that these parameters themselves are dynamical, being related to vacuum
expectation values of various fields. The ultimate dream of a string
theorist is that when we fully understand the dynamics of string theory,
we shall find some dynamical principle which determines a unique point in
this parameter space, and that this point will correctly describe the
nature that we see.

\subsection{D-branes} \label{s2.2}

Dirichlet(D)-branes are soliton like configurations in type IIA/IIB/I
string
theories\cite{LEIGH,9510017}. However, the description of D-branes is
quite different from the
way we normally describe a conventional soliton. To understand this
distinction, let us review the way a conventional soliton is described in
a quantum field theory:
\begin{enumerate}

\item First we construct a time independent solution of the classical
equations of motion
with
energy density localised around a $p$-dimensional spatial hypersurface.
We shall call such a solution a $p$-brane soliton. Thus for example
a 0-brane corresponds to a particle like soliton,
a 1-brane corresponds to a string like soliton, a 2-brane
corresponds to a membrane like soliton and so on.

\item Next we identify the fluctuation modes of various fields around this
background solution which are {\it localised} around the brane.
These modes describe the fluctuation of the brane around its equilibrium
position, and include for example
oscillation modes of the brane in directions transverse to the brane.

\item We then construct the $(p+1)$ dimensional field theory which
describes 
the dynamics of these modes.
(This can be derived from the original field theory whose
soliton we are analysing.) This field theory describes the classical
dynamics of the soliton solution.

\item In order to study the quantum dynamics of this soliton, we quantize
this field theory.
\end{enumerate}
In describing a D-brane we start from step 4 and work backwards.

\begin{figure}
\begin{center}
\leavevmode
\epsfbox{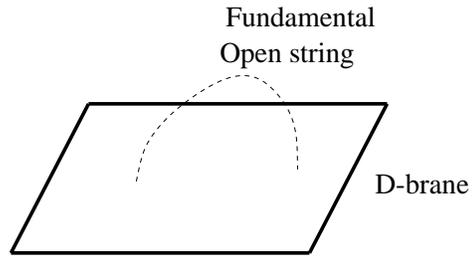}
\end{center}
\caption{Open string excitations on a D-brane.} \label{f6}
\end{figure}

I shall now give explicit description of D-branes in type IIA and type
IIB string theories.  
Elementary excitations in type IIA / IIB string theories are {\it closed
strings}.
D-$p$-branes in these theories are $p$-dimensional soliton like objects
whose quantum
dynamics  
is described by the theory of {\it open strings} whose
ends are constrained to move on the D-brane. This has been illustrated in
Fig.\ref{f6}. Starting from this description we can work backwards
and derive all the properties of these branes. It turns out that 
type IIB string theory has stable (BPS) D-$p$-branes for $p$-odd (1, 3,
5, 7,
9), whereas
type IIA string theory has stable D-$p$-branes for $p$-even (0, 2,
4, 6,
8).
In particular, type IIB string theory contains a D9-brane which
fills all space. For this theory one can also add to this list the $p=-1$
brane, representing a configuration that is localised not only in all
space directions but also in the
time direction. This is called the D-instanton.

Under duality often elementary closed string states in one theory get
mapped to
D-brane states in the dual theory.
Indeed, 
this is how the importance of D-branes in the study of non-perturbative
string
theory was first realised\cite{9510017}.
But since their discovery D-branes have found many applications in string
theory.

A particularly important property of D-branes is as follows\cite{9510135}.
{\it The $(p+1)$ dimensional effective field theory, describing the
dynamics of $N$ coincident D-$p$-branes at low energy, is a 
supersymmetric
U(N) gauge theory with 16 supercharges.} This is an ordinary
gauge theory with a set of Majorana fermions and scalars in the
adjoint
representation of the gauge group, with the Yukawa and scalar
self-couplings determined in terms of the gauge coupling via specific
relations. We shall encounter the (3+1)-dimensional version of this theory
again later while discussing the Maldacena conjecture.

\subsection{Black Hole Entropy and Hawking Radiation} \label{s2.3}

Black holes are classical solutions of general relativity.
They can be formed from collapse of matter under gravitational
pull.
Classically a black hole is completely black, {\it i.e.}
it absorbs everything that falls in without emitting anything.
But this picture changes significantly in the quantum theory.
It turns out that in the quantum theory, black holes behave as perfect
black bodies
at {\it finite temperature} proportional to the inverse mass of the black 
hole. In particular,
\begin{itemize}
\item black holes emit thermal radiation known as Hawking radiation, and
\item a black hole carries an entropy proportional to its surface area,
known as
the 
Bekenstein-Hawking entropy.
\end{itemize}

These results were derived using a semi-classical analysis, but there was
no microscopic (statistical) understanding of this entropy and radiation
in this semi-classical treatment. In particular, the expressions for the
entropy and the Hawking temperature were given in terms of classical
geometrical properties of the space-time around the black hole. It was as
if we had themodynamics without statistical mechanics. This led Hawking to
suggest that starting with a collapsing spherical shell of matter in a
pure quantum state, and letting it collapse into a black hole and
subsequently evaporate via Hawking radiation, we can have a pure quantum
state get transformed to a mixed state described by a thermal density
matrix. Such a process violates the laws of quantum mechanics. 

It turns out that in string theory, for a special class of black holes,
one can find dual
descriptions as:
\begin{itemize}
\item a classical solution of the equations of motion, and
\item a configuration of D-branes.
\end{itemize}
Thus for these black holes we can compute entropy and Hawking radiation in
two ways:
\begin{itemize}
\item Use Bekenstein and Hawking's formul\ae\ to compute the entropy and
the
rate of radiation from the classical solution.
\item Use the quantum string theory living on the D-brane to compute the
number of quantum states $N$ and the rate of radiation from the system.
From this we can compute the entropy by taking the logarithm of the
degeneracy $N$.
\end{itemize}

These two computations give identical
result\cite{9601029,9602043,9605234,9606185,9609026}.\footnote{The two
computations are valid in
different regions of the parameter space, and we need to continue the
result computed in one region to the other region by invoking
supersymmetric non-renormalization theorems.} Thus
this analysis provides a microscopic explanation of the black hole entropy
and
Hawking radiation for these special class of black holes.
It remains to be seen whether these results generalize to the
more general black holes.

\sectiono{String Theory: 1997-2000} \label{s3}

The developments in string theory during this period can be described as
study of various aspects of $U$-theory. I shall discuss some of them
here.

\subsection{Matrix Theory} \label{s3.1}

Note that one corner of the parameter space of $U$-theory is $M$-theory. 
$M$-theory is known to reduce to 11-dimensional supergravity theory at low
energy.  But there is no known systematic procedure for computing
corrections to this low energy approximation.  {\it Matrix
theory\cite{9610043} is a
proposal for defining $M$-theory beyond this low energy approximation.}
Although I shall not describe the logic which led to the formulation of
Matrix-theory, let me just state that this proposal comes from examining
the
dynamics of D0-branes in IIA in the appropriate limit in which it
approaches
the $M$-theory corner\cite{9608024,9709220,9710009}.

The Matrix theory proposal is as follows.  Consider a scattering
process in $M$-theory:  $A + B\to C + D +\ldots$ for a set of external
states $A$, $B$, $C$, $D$, $\ldots$. Let us go to a frame in which the
centre of mass has an infinite boost.  This is known as the infinite
momentum frame.  {\it
According to Matrix theory, $M$-theory in the infinite momentum frame is
equivalent to a quantum mechanical system.} The dynamical variables in
this quantum mechanical system are $N\times N$ matrices where $N$ is
proportional to the total momentum of the original system.  Since in the
infinite
momentum frame the total momentum of any system is infinite, we need to
take the $N\to\infty$ limit at the end. 

This, in principle, gives an algorithm for computing any physical
quantity in $M$-theory by mapping it to an appropriate quantity in this
quantum mechanical system. We can perform various
consistency checks. In particular at low energy, scattering amplitudes
computed
using matrix quantum mechanics must agree with those computed
from 11-dimensional supergravity.
Matrix theory has passed many such tests\cite{9705091,9706072}.
Of course, 
in principle one should be able to use matrix theory to go beyond
tree level supergravity.
This has not been achieved so far.

\subsection{Maldacena Conjecture} \label{s3.2}

The starting point here is the study of $N$ coincident D3-branes in
type IIB string theory in
the large $N$ limit.
In this limit the system has dual descriptions
\begin{itemize}
\item
as a solution of the classical equations of motion of string theory /
supergravity, and
\item as a D-brane system.
\end{itemize}
Requiring that these two descriptions are equivalent led Maldacena to the
following conjecture\cite{9711200}:
{\it Type IIB string theory on $(AdS)_5\times S^5$ is equivalent to
$\NN=4$ supersymmetric SU(N) gauge theory in (3+1) dimensions.}

There are several terms in the above statement which may not be familiar,
so I shall now define them.
$S^5$ is an ordinary five dimensional sphere of radius $R$ defined by the
equation
\be \label{e3.1}
\sum_{i=1}^6 (y_i)^2 = R^2\, ,
\ee
where $y_i$ are the coordinates of a six dimensional Euclidean space.
$(AdS)_5$ is the five dimensional anti-de Sitter space described by the
equation
\be \label{e3.2}
(x_0)^2 +(x_1)^2 - (x_2)^2 - (x_3)^2 -(x_4)^2 -(x_5)^2 =
R^2,
\ee
where $x_i$'s describe a six dimensional space with metric
\ben \label{e3.3}
ds^2 &=& -(dx_0)^2 -(dx_1)^2 + (dx_2)^2 + (dx_3)^2 \nonumber \\
&& +(dx_4)^2
+(dx_5)^2 \,
.
\een
It turns out that the
boundary of $(AdS)_5$ is a (3+1) dimensional Minkowski space. Often in
order to make precise statements one needs to Wick rotate to the Euclidean
version of $AdS_5$. This is obtained by changing the signs of the
$(x_1)^2$ term in eq.\refb{e3.2} and the $(dx_1)^2$ term in
eq.\refb{e3.3}. Its boundary is a four dimensional Euclidean space.

Finally we need to define
$\NN=4$ supersymmetric SU(N) gauge theory. This is  an ordinary SU(N)
gauge theory
with 6
scalars and 4 Majorana fermions in the adjoint representation of
the gauge group, with specific relations between the gauge coupling
constant and various Yukawa
and scalar self-couplings. In particular all the Yukawa and scalar
self-couplings are determined in terms of the gauge coupling. This theory
turns out to be a conformally invariant (3+1) dimensional field theory,
{\it i.e.} the $\beta$-functions in this theory vanish.

According to the Maldacena conjecture,
the relation between the dimensionless parameters of IIB string theory on
$AdS_5\times
S^5$ and supersymmetric $SU(N)$ Yang-Mills theory is as follows:
\be \label{e3.4}
g_{string} = g_{YM}^2, \qquad R= (4\pi g_{YM}^2
N)^{1/4}\, ,
\ee
where $g_{YM}$ and $g_{string}$ denote the coupling constants of the
$\NN=4$ supersymmetric Yang-Mills theory and the type IIB string theory
respectively, $R$ is the radius of $S^5$ and $AdS_5$ as introduced
through eqs.\refb{e3.1} and \refb{e3.2}, and $N$ is the $N$
of $SU(N)$. 

We are now in a position to state the precise form of the Maldacena
conjecture\cite{9802109,9802150}:
{\it There is a one to one correspondence between the physical Greens
functions in type IIB string theory on $(AdS_5\times S^5)$ and
the correlation functions of gauge invariant operators in $\NN=4$
supersymmetric SU(N) gauge theory on the boundary of $(AdS)_5$, with the 
identification of the parameters as given in eq.}\refb{e3.4}.

Consider now the 't Hooft large $N$ limit\cite{LARGEN}:
\be \label{e3.5}
g_{YM}\to 0, \quad N\to\infty, \quad \lambda\equiv
g_{YM}^2
N \, \, {\rm fixed}\, .
\ee
This gives, using eq.\refb{e3.4}
\be \label{e3.6}
g_{string}\to 0, \qquad R= (4\pi \lambda)^{1/4}\,
\, {\rm fixed}\, .
\ee
Thus
if $\lambda$ is large then $R$ is large. In this limit the smallness of
$g_{string}$ implies that we can ignore the string loop corrections, {\it
i.e.} restrict ourselves to string tree diagrams or classical string
theory. The largeness of $R$  on the other hand allows us to use the low
energy approximation to string theory, which is type IIB supergravity
theory. Thus  in this case, quantum string theory can be approximated
by classical supergravity.

Many other examples of this kind of relationship have been found by
studying other brane systems\cite{9711200}.
The generic form that such a relation takes is that 
a string theory / $M$-theory on a manifold $K$
is equivalent to a quantum field theory on the boundary of $K$.
The precise form of this quantum field theory depends on the
choice of $K$ and the particular string theory / $M$-theory that we are
using.

There have been various applications of Maldacena conjecture, both for
using supergravity / string theory to study strong coupling limit of gauge
theories, and using gauge theory to study non-perturbative aspects of
string theory. Here I shall discuss one application of each kind. 
It had been conjectured earlier that {\it in a consistent quantum theory
of gravity,
the fundamental degrees of freedom reside at the boundary of
space-time and not in the interior.
Furthermore, there is
$\sim 1$ degree of freedom per Planck
`area'.}\footnote{Note that if the boundary is an $n$-dimensional space,
then the area refers to the $n$-volume of this space.} This principle is
known as the holographic principle\cite{9310006,9409089}.
Maldacena
conjecture provides a concrete verification
of the holographic principle for type IIB string theory on $(AdS)_5\times
S^5$ by
relating string theory on $AdS_5\times
S^5$ to a gauge theory living on its boundary. In the
$\NN=4$ supersymmetric
gauge theory described earlier, it is
straightforward to
count the total number of degrees of freedom after a suitable ultra-violet
regularisation of the gauge theory. It turns out that this
ultra-violet cut-off is related to the infrared cut-off of the
string theory on $AdS_5\times S^5$, so that with this cut-off in place, 
the
boundary of $AdS_5$ at a given instant of time, instead of having infinite
volume like an
ordinary 3 dimensional space, has finite volume. Taking the
ratio of the number of degrees of freedom of the gauge theory and
the volume of this 3-dimensional space, we find that the cut-off
dependence goes away, and there is indeed
one (up to a numerical factor) degree of freedom per Planck volume living
on the boundary of
$AdS_5$\cite{9805114}.

\begin{figure}
\begin{center}
\leavevmode
\epsfbox{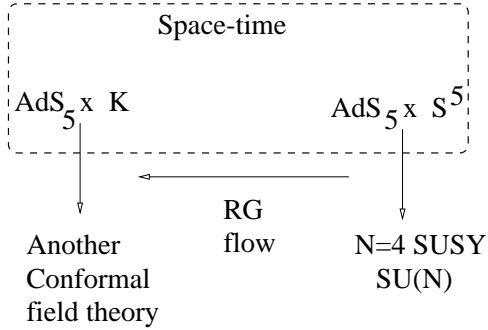}
\end{center}
\caption{Renormalization group flow and solutions of supergravity
equations of motion. The upper part of the diagram shows the space-time
picture of the supergravity solution. To the far right (the UV region) the
space-time looks like a piece of $AdS_5\times S^5$, whereas to the far
left (IR
region) the space-time looks like a piece of $AdS_5\times K$ for some
manifold $K$. The lower part shows
the interpretation of different regions of the supergravity solution as
quantum field
theories.} \label{f7} 
\end{figure}

Another application of the Maldacena conjecture has been in the study of
renormalization group (RG) flows in conformal field
theories\cite{9810126,9810206,9904017}.
For example, we can perturb the $\NN=4$ superconformal field theory by
adding some
relevant perturbation.
The resulting theory either
\begin{itemize}
\item
flows to a conformal field theory, or
\item flows to a theory with a mass gap ({\it e.g.} a confining theory).
\end{itemize}
On the supergravity side this perturbed theory should be described by a
solution which in some region of space-time (UV) looks like $AdS_5\times
S^5$ background, but in another region (IR) differs from the $AdS_5\times
S^5$ geometry and represents the infra-red theory. The situation has been
illustrated in Fig.\ref{f7}.
Thus the RG evolution parameter becomes a coordinate in space time.
For all RG flows which can be described via the supergravity
configurations of this type, the $c$-theorem (which states that the
central charge of a conformal field theory decreases along the RG flow) 
follows
naturally from general properties of the supergravity equations of motion.
This procedure has also been used to describe RG flow to theories with
mass gap and
confinement, where in the infrared the theory does not flow to a
conformal field theory\cite{0003136}.
In these cases space-time at the far left end (representing infrared
direction
in RG flow) has a more complicated structure\cite{0003136}.

\subsection{D-branes and Non-commutative Field Theories} \label{s3.3}

The dynamics of a D-$p$-brane is described by an open string theory,
which could be regarded as a $(p+1)$-dimensional field
theory with infinite number of fields. In the low energy limit only a
finite number of massless fields survive whose dynamics is described by
a $(p+1)$-dimensional effective
quantum field theory. One of these fields is a U(1) gauge field.
We can study this $(p+1)$-dimensional field theory in the presence of
constant background electric / magnetic field associated to this U(1)
gauge field. It turns out that if we switch on a constant background
magnetic field, and take a suitable limit involving the string tension,
the string coupling constant, the background metric and the background
magnetic
field, then the dynamics
of the D-brane in this limit is described by a non-commutative gauge 
theory\cite{9711162,9711165,9903205,9908142}. The
action for this non-commutative gauge theory is obtained by starting with
the original low energy effective action that we had for zero magnetic
field background, and then  {\it
replacing all products of fields in
the original action by the non-commutative product:}
\ben \label{e3.8}
A(x) B(x) &\to& A(x) * B(x) \nonumber \\
&=& \exp(\Theta^{\mu
\nu}\partial_\mu\partial'_\nu) A(x) B(x')|_{x=x'}\, , \nonumber \\
\een
where
$\Theta^{\mu\nu}$ is an anti-symmetric matrix determined in terms of the
strength of the background
magnetic field and other parameters.
Thus one can use non-commutative field theory results to study D-branes in
the presence of background magnetic field and vice-versa.

One of the important results coming out of these studies is that
{\it the infrared and ultraviolet effects do not decouple in a
non-commutative
field theory\cite{9912072}.}
Thus physics at large distance can affect physics at short distance and
vice versa. There have been suggestions that this is also a general
property
of string theory\cite{0007146}. Another important result is the discovery
that non-commutative scalar field theories can contain stable soliton
solutions\cite{0003160} even if their commutative counterparts do not
contain such solutions due to Derrick's theorem\cite{DERRICK}.

\subsection{Tachyon Condensation on Brane-Antibrane System} \label{s3.4}

As has been discussed earlier,
type IIB string theory
has stable D$p$-branes for odd $p$.
In particular,
$p=9$ describes a space-filling 9-brane.
This theory also has stable D$\bar p$-branes for odd $p$.
These are
D$p$-branes with opposite orientation.
It turns out that
although D$p$-branes and D$\bar p$-branes are individually stable, a
system
of coincident D$p$-brane D$\bar p$-brane has tachyonic modes.
These are scalar fields $T$ with
negative mass$^2$.
Thus such a system is
classically unstable.

The question that arises naturally is:
Is there a stable minimum of the {\it classical} tachyon potential
$V(T)$?
The conjectured answer to this question is as
follows\cite{9904207}:
\begin{enumerate}
\item The minimum of $V(T)$, $T=T_0$, describes `nothing' where the
original
energy of the brane-antibrane system is exactly cancelled by $V(T_0)$
\item There are no perturbative open string excitations around this
vacuum, but there are solitons.
These solitons describe lower dimensional D-branes.
\end{enumerate}

Evidence for these conjectures come from various sources (string field
theory\cite{9912249,0001084,0002237,0005036,0009103,0009148,0009191},
conformal
field
theory\cite{9808141,0003101}, non-commutative field
theory\cite{0003160,0005006,0005031}, Maldacena conjecture\cite{0004131},
Matrix theory\cite{0010016,0010058,0011094} etc.) There are several
applications of these results:
\begin{itemize}
\item Describing all D-branes as solitons on space-filling branes gives a
way to classify all possible D-brane configurations in a string theory in
terms of {\it K-theory}\cite{9810188,9812135}.
\item These results suggest that the open string theory on the
space-filling brane anti-brane
system may provide a non-perturbative formulation of string theory, since
\begin{itemize}
\item it naturally contains all D-branes as solitons, and
\item there is also evidence that this theory contains closed string
states\cite{0005031,9901159,0002223,0009061,0010240}.
\end{itemize}
{}From this viewpoint, the fundamental degrees of freedom are open
strings,
and closed strings (including gravitons) and D-branes are composite
objects.
\end{itemize}

\subsection{Special Limits of String Theory: String Theories without
Gravity}
\label{s3.5}

If we have a $p$-brane soliton, then the degrees of freedom on
the soliton describe a quantum theory living in $(p+1)$ dimensions.
Typically these degrees of freedom interact with those living in the bulk, 
and we do not have a consistent quantum theory involving only the degrees
of freedom living on the $p$-brane.
But in certain special limits, the degrees of freedom in the bulk may
decouple.
In that case, we get 
a consistent $(p+1)$-dimensional quantum theory on the $p$-brane
world-volume {\it without gravity}  since gravity lives
in the bulk. One such limit is the low energy
limit
in which we recover a quantum field theory living on the
$p$-brane
world-volume. As discussed in subsection \ref{s3.3}, a
variation of
this limit can also give rise to non-commutative field theories. But there
are other limits which give us full fledged string
theory (or other quantum theories which cannot be described as quantum
field theories) in $(p+1)$-dimensions without gravity. These theories are
expected to capture many of the important features of standard string
theories, without being plagued by the conceptual problems which arise due
to the presence of gravity. I shall give a few examples here.

\begin{itemize}

\item LST (Little string theories): Besides containing the D-brane
solitons, string theories also contain another kind of 5-brane soliton,
known as the NS 5-brane. In the presence of a set of coincident NS
5-branes, we have a
set of degrees of freedom localised on the 5-branes, describing the
dynamics of these NS 5-branes, and
another set of degrees of freedom living in the (9+1) dimensional bulk
space-time. It turns out that in the limit of vanishing
string coupling constant with the string tension remaining finite,
the degrees
of freedom living on the NS 5-branes decouple from the
degrees of freedom
living in the bulk, and hence we get a consistent (5+1) dimensional string
theory without gravity\cite{9705221,9707250,9911147}. This theory and
its various cousins obtained via compactification are known as little
string theories.

\item NCOS (Non-commutative open string) theory: In this case the starting
point
is a D-$p$-brane in the presence of a non-zero constant electric field
background. By taking an appropriate limit involving the string coupling
constant, the string tension, the background metric and the electric
field strength on the
D-$p$-brane one can decouple the bulk modes from the modes living on the
D-$p$-brane\cite{0005040,0005048}. The $(p+1)$ dimensional theory obtained
this way
has been called
non-commutative open string theories, since the fundamental degrees of
freedom in this theory are open strings, but the action involving these
open strings is related to the action of the usual open string theory by a
replacement of all ordinary products by appropriate non-commutative
products.

\item OM (open membrane) theory: Like string theory, $M$-theory also has 
solitonic branes. In particular, it has a 5-brane and a 2-brane soliton
solutions. Among the degrees of freedom living on the 5-brane
world-volume, there is a rank 2 anti-symmetric tensor gauge field. Its
field strength is a totally anti-symmetric self-dual rank 3 tensor in
(5+1)
dimensions. It turns out that starting from a configuration with constant
non-vanishing field strength of this anti-symmetric tensor field, and
taking an appropriate limit involving the Planck mass,
the background metric
and this field strength, one can decouple the bulk modes from the brane
modes\cite{0006062}. The resulting (5+1) dimensional quantum theory living
on the
5-brane has been called the open membrane theory. As the name suggests,
excitations in this theory include
open membranes with their
boundaries stuck to the 5-brane.

\item OD$p$ (Open D$p$-brane) theories: As in the case of little string
theories, here the starting point is an NS 5-brane of type IIA string
theory. But instead of putting it in trivial background space-time, we now
switch on constant background values of some appropriate bulk fields
(known as Ramond-Ramond (RR) gauge fields). By taking appropriate limits
of the string tension, the string coupling constant, the background metric
and the background value of the RR gauge fields, one can again decouple
the theory in the bulk from the theory on the brane\cite{0006062}. The
result is a quantum theory on the 5-brane, known as open D-$p$-brane
theories. The name originates from the fact that the excitations on the
brane include open D-$p$-branes (the value of $p$ depends on which
particular RR gauge field is switched on) with their boundaries stuck on
the
5-brane. 

\end{itemize}

\subsection{Large Radius Compactification} \label{s3.6}

In conventional compactification of string theory leading to
semi-realistic models, both gravity and gauge fields come from the closed
string sector.
In this scheme, a direct
upper bound on the size of the compact dimension comes from experimental
verification of various force laws to small distance scales.
Thus, for example,
test of QED down to a distance scale of $(TeV)^{-1}$ will mean
that the size of the compact direction cannot be larger than $(TeV)^{-1}$.
Otherwise we would have to use higher dimensional QED for our computation.

\begin{figure}
\begin{center}
\leavevmode
\epsfbox{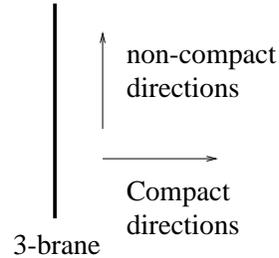}
\end{center}
\caption{Compactification with branes.} \label{f8}
\end{figure}

Discovery that gauge fields can live on D-branes has given rise to new
possibilities\cite{9602070,9603133,9606036,9803315,9804398}.
Consider for example the scenario shown in Fig.\ref{f8} where 6 of the
directions are
compactified on a manifold $K$, and there is a set of three branes
with their world-volume directed along the non-compact directions,
situated at a given point in the
compact space.
All gauge fields and known matter fields come from the world-volume theory
on the three brane, and gravity comes from the closed strings living
in the bulk. Thus here
gauge fields are always (3+1) dimensional, irrespective of the size of the
compact dimensions. 
In this scenario, the only direct upper bound on the size of the compact
dimension comes from the test of inverse square law for gravity at short
distance scale.
This gives an upper bound 
of about a millimetre on the size of the compact dimensions. 
There are of course other indirect bounds which we shall not discuss here,
but the fact remains that the size of these extra compact dimensions can
be much larger than $(TeV)^{-1}$, $-$ the distance scale probed in the
present
accelerator experiments.

Presence of large extra dimensions also changes the relation between four
dimensional Planck scale and string scale. The new relation is of the
form:
\be \label{e3.21}
M_{Planck} \sim M_{string} \sqrt{M_{string}^6
V_{compact}}\, ,
\ee
where $M_{Planck}$ denotes the four dimensional Planck mass, $M_{string}$
denotes the square root of the string tension, and $V_{compact}$ denotes
the volume of the
compact six dimensional manifold. (We have taken the string coupling to be
of order 1). From this relation we see that if the size of the compact
direction is much larger than the string scale, {\it i.e.} if
$M_{string}(V_{compact})^{1/6}>>1$, then $M_{string} <<
M_{Planck}$.
It has even been suggested that 
$M_{string}$ can be of the order of  a TeV; this requires
$M_{string}(V_{compact})^{1/6}$ to be of order $10^5-10^6$. This would
resolve the usual
hierarchy problem as to why $M_{string}$ is so large compared to
the mass scale of weak interaction $-$ in this scheme they are of the
same order. However, it gives rise to a new
hierarchy problem, $-$ that of explaining why
$M_{string} (V_{compact})^{1/6}$ is so large. Various ideas have been put
forth,
but there is no definite conclusion.

\begin{figure}
\begin{center}
\leavevmode
\epsfbox{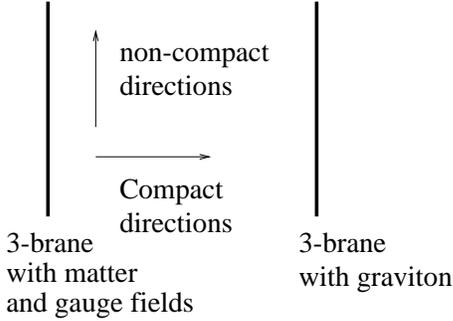}
\end{center}
\caption{Illustration of the Randall-Sundrum scenario.} \label{f10}
\end{figure}
A twist to this tale is provided by the 
Randall-Sundrum scenario for string
compactification\cite{9905221,9906064} illustrated in Fig.\ref{f10}. In
this
scenario,
the graviton that we observe in four dimension is not the usual ten
dimensional bulk graviton carrying zero momentum in the compact direction,
but a particular mode of the ten dimensional graviton which is localised
on a brane. This
brane however is not the brane that we live on, but another brane (which
we shall call the gravity brane),
separated from us by a fairly large distance along the compact direction.
Since 
the wave-function of the graviton falls off exponentially away from the
gravity 3-brane.
it
has a very small value on the matter 3-brane.
This fact can be used to explain why gravity couples so weakly to matter,
{\it i.e.} why the
effective Planck mass in our (3+1) dimensional world is so large. The
advantage of the scenario is that due to the exponential fall off of the
graviton wave-function away from the gravity brane, one can generate a
large hierarchy between the effective four dimensional Planck scale and
the weak scale without having to make the actual
separation between the gravity brane and our brane very large.

\sectiono{Summary} \label{s4}

I shall end by summarising the main points once more.
\begin{itemize}
\item
String theory has had reasonable success in providing a
consistent quantum theory of gravity. This includes:
\begin{enumerate}
\item Finiteness of perturbation theory
\item Partial resolution of the problems associated with quantum
mechanics of black holes
\item Explicit realization of holographic principle in certain
backgrounds
\end{enumerate}

\item
String theory also has the potential for providing a unified
theory of all interactions. In particular, it can give rise to
\begin{enumerate}
\item Gauge group containing $SU(3)\times SU(2)\times U(1)$
\item Chiral fermions
\item Three generations of quarks and leptons
\item N=1 supersymmetry
\end{enumerate}

\item It has also proved to be an internally consistent and
beautiful theory.
In particular, string duality provides

\begin{enumerate}
\item Unification of all string theories.
\item Democracy between elementary and composite particles.
\item Unification of classical and quantum effects.
\end{enumerate}

\item Progress in string theory has also dramatically improved
our understanding of various aspects of quantum field theories and
other quantum theories based on extended objects. New relationship between 
quantum field theories and string theories have been discovered during the
past few years.

\item The last few years have seen several attempts at giving a
non-perturbative definition of string/M(U) theory. It is still too early
to say if any of them will give rise to a fruitful approach to the study
of
non-perturbative effects in string theory.

\item The brane world scenario has given rise to novel possibilities for
string compactification.

\end{itemize}

However, despite this enormous progress, string theory
is still
far from achieving its final goal, which is to provide a unified
theory of all matter and their interactions.

{\noindent \bf Acknowledgment}:
I wish to thank S.~Rao for comments on the manuscript.

\end{document}